\documentclass[10pt,sigconf,letterpaper]{acmart}

\renewcommand\footnotetextcopyrightpermission[1]{} %
\setcopyright{none}
\usepackage{color}
\usepackage{pifont}
\usepackage{xcolor}

\iftrue

\newcommand{\geoff}[1]{\ding{110}\ding{43}\textcolor{violet}{#1}}
\else

\newcommand{\geoff}[1]{}
\fi
\newcommand{\ids}{UID smuggling}
\newcommand{\IDS}{UID Smuggling}

\newcommand{\dsmugglers}{dedicated smugglers}

\newcommand{\dsmuggler}{dedicated smuggler}

\newcommand{\Web}{Web}

\copyrightyear{2022}
\acmYear{2022}
\acmConference[In submission]{}{conference TBD}{date TBD}
\acmBooktitle{Internet Measurement Conference (IMC '22), October 25--27, 2022, 
Nice, France}

\begin{document}

\author{Audrey Randall}
\email{aurandal@eng.ucsd.edu}
\affiliation{UC San Diego}
\author{Peter Snyder}
\email{pes@brave.com}
\affiliation{Brave Software}
\author{Alisha Ukani}
\email{aukani@ucsd.edu}
\affiliation{UC San Diego}
\author{Alex Snoeren}
\email{snoeren@cs.ucsd.edu}
\affiliation{UC San Diego}
\author{Geoffrey M.\ Voelker}
\email{voelker@cs.ucsd.edu}
\affiliation{UC San Diego}
\author{Stefan Savage}
\email{savage@cs.ucsd.edu}
\affiliation{UC San Diego}
\author{Aaron Schulman}
\email{schulman@cs.ucsd.edu}
\affiliation{UC San Diego}


\title{Measuring UID Smuggling in the Wild}
\renewcommand{\shortauthors}{A. Randall \textit{et al.}}



\begin{abstract}
This work presents a systematic study of \ids{}, an emerging tracking 
technique that is designed to evade browsers' privacy protections. Browsers are 
increasingly attempting to prevent cross-site tracking by partitioning the 
storage where trackers store user identifiers (UIDs). \ids{} allows trackers to 
synchronize UIDs across sites by inserting UIDs into users' navigation 
requests. Trackers can thus regain the ability to aggregate users' activities 
and behaviors across sites, in defiance of browser protections. 
%
%

In this work, we introduce CrumbCruncher, a system for measuring UID smuggling 
in the wild by crawling the Web. CrumbCruncher provides several improvements 
over prior work on identifying UIDs and measuring tracking via web crawling, 
including in distinguishing UIDs from session IDs, handling dynamic web 
content, and synchronizing multiple crawlers. We use CrumbCruncher to measure 
the frequency of \ids{} on the Web, and find that 
\ids{} is present on more than eight percent of all navigations that we made. 
Furthermore, we perform an analysis of the entities involved in \ids{}, and 
discuss their methods and possible motivations. 
%
%
We discuss how our findings can be used to protect users from \ids{}, and 
release both our complete dataset and our measurement pipeline to aid in 
protection efforts.
\end{abstract}

\maketitle

\section{Introduction}
\label{sec:intro}

Over the past few years, tensions have deepened between those
collecting detailed user behavior data for advertising purposes and
privacy-conscious users who do not want to be monitored.  While there are
some efforts to find a compromise between these positions (e.g.,
allowing the collection of aggregated, anonymized
data~\cite{google_blog_floc, google_blog_topics}), none have yet
managed to satisfy advertisers or privacy
advocates~\cite{eff_floc,wsj_floc, bbc_floc}. In the absence of such a
solution, privacy-focused browsers (i.e., browsers for which privacy
is seen as a competitive advantage) have rolled out changes that block
one of the core mechanisms used by \emph{third-party trackers} to aggregate
information about a user \emph{across different websites}, thereby
building a profile of that user.  

Previously, third-party trackers could build user profiles across websites 
because information stored by the tracker was accessible to that tracker across 
all websites that include it.
Trackers commonly used third-party cookies
for this purpose, although any type of browser storage could be used.
Trackers could use this shared storage to
build shared state for each user across every website that included
the tracker.  However, several browsers are now employing an 
anti-tracking defense called ``partitioned storage,'' which removes this sharing
ability. By partitioning all browser storage by the domain of the top-level
website, browsers intended to prevent trackers from linking user information 
across sites.

However, trackers have responded by implementing a new class 
of tracking technique that we call \emph{\ids{}}. \ids{} allows trackers to share a user's 
information across websites by modifying the user's navigation requests. The 
tracker accomplishes this style of tracking by decorating users' navigation 
requests with identifying information, which will then be shared across first-party boundaries. The tracker may also choose to momentarily redirect the user 
to its own domain, where it can record this smuggled information as a first 
party itself. In each case, trackers use \ids{} to regain the ability to link  
user-identifying information across sites, circumventing the browser's attempt 
to partition such information.

This work presents the first systematic measurement of \ids{} in the wild.
We make the following contributions to understanding online tracking and 
improving Web privacy:

\begin{enumerate}
	\item We perform the \textbf{first systematic measurement} of \ids{}
	in the wild.
	\item We construct a multi-stage \textbf{analysis pipeline}, nicknamed 
	CrumbCruncher, to crawl the \Web{} and measure how frequently \ids{} occurs.
	\item We improve on prior techniques for \textbf{differentiating user 
	identifiers} from other values and \textbf{synchronizing multiple crawlers}.
	\item We categorize the \textbf{behaviors} of trackers, including which 
	categories of sites are more likely to engage in \ids{}.
	\item We contribute to countermeasures against \ids{}, both by 
	sharing our hand-edited \textbf{dataset}, and 
	by publishing our tool for finding new instances, CrumbCruncher.
\end{enumerate}

The remainder of our work is organized as follows. 
Section~\ref{sec:background_related} covers the background of navigational 
tracking and related work. Section~\ref{sec:methodology} describes 
the design of our crawler, CrumbCruncher, and its capabilities and limitations. 
Section~\ref{sec:results} presents 
our findings, including the most common participants in navigational tracking 
and a summary of their behaviors and categories. Section~\ref{sec:limitations} 
describes the limitations of our work. 
Section~\ref{sec:countermeasures} details our contribution to the 
countermeasures that various entities have taken against navigational tracking. 
Section~\ref{sec:related} discusses related work, 
and Section~\ref{sec:conclusion} 
concludes.


%

\section{Background}
\label{sec:background_related}

Advertisers want to track user activity across sites for a variety of
purposes, including performing identity resolution and supporting
affiliate marketing, but such capabilities represent a significant
threat to user privacy.

For over a decade, browsers allowed advertisers to perform these
cross-site tracking functions with third-party cookies. But because
this capability presents a threat to privacy, several
popular browsers have implemented \emph{partitioned storage} to isolate
third-party cookies so they cannot be used for cross-site tracking.
At the time of writing, Firefox, Safari, and 
Brave~\cite{brave_partitioned_storage,
  safari_partitioned_storage, firefox_partitioned_storage} all use
partitioned storage by default.

\begin{figure}[t]
	\centering
	\includegraphics[width=3in]{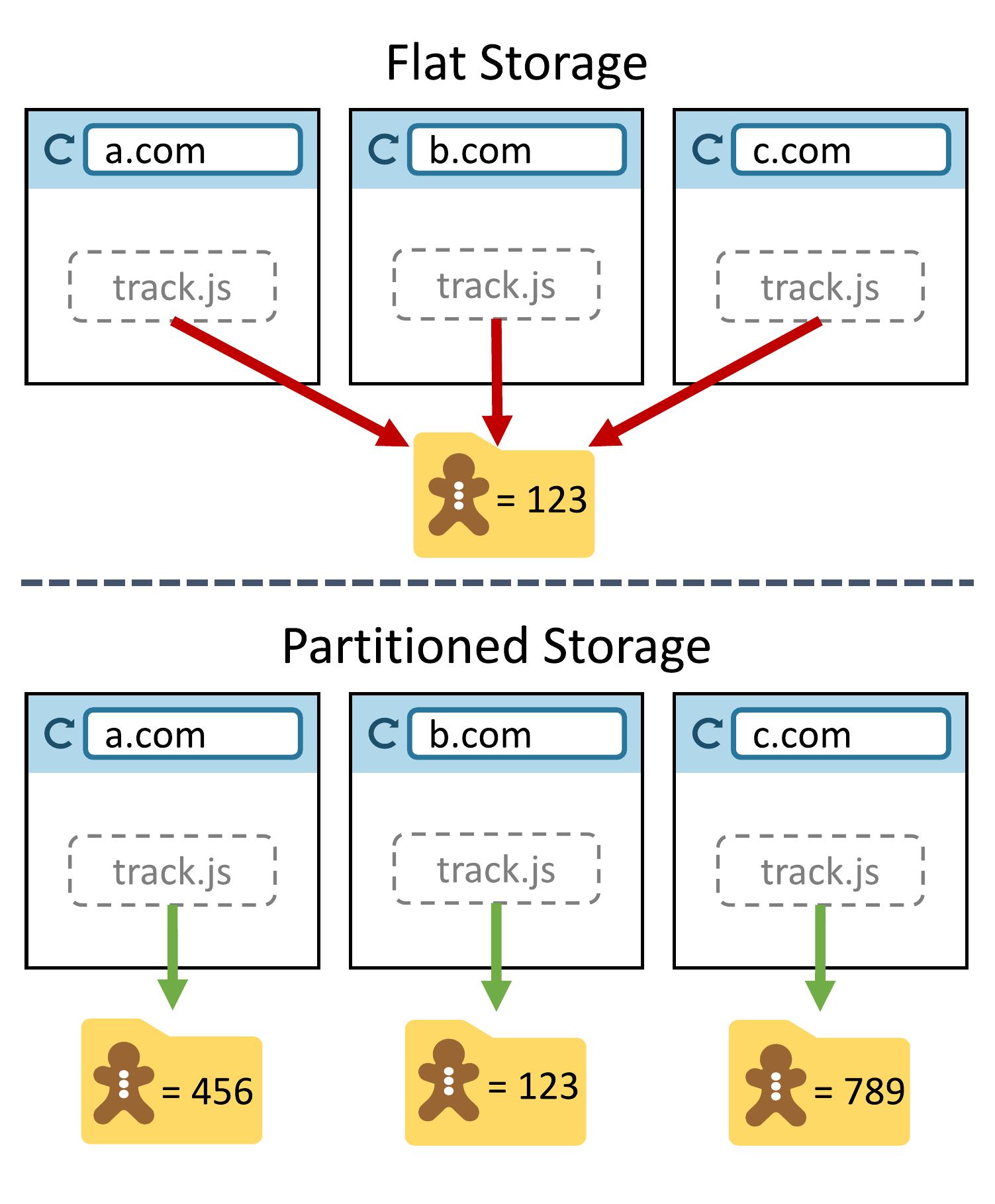}
	\caption{Flat storage versus partitioned storage.}
	\label{fig:hierarchical_structure}
\end{figure}

Partitioned storage uses a hierarchical namespace, where the hierarchy
is based on the domain of the frame that contains the cookie-storing element. 
Figure~\ref{fig:hierarchical_structure} shows the difference between 
flat and partitioned storage from a tracker's perspective. When flat storage is 
in use, the tracker can read from or write to the same storage area regardless 
of which website it is on, but when partitioned storage is implemented, the 
tracker accesses a different storage area on each website that loads the 
tracker. This prevents trackers from assigning the same user-identifying cookie 
(represented by the gingerbread man icon) to users across sites. A similar 
system is used for other browser storage, such as local storage.

To circumvent the protections that partitioned storage provides,
advertisers are increasingly using \emph{\ids{}}.
\ids{} modifies a user's navigation requests by adding information
to the navigation URLs in the form of query parameters. \ids{} may also redirect
the user to one or more third-party trackers before redirecting to the intended 
destination.

\begin{figure*}
	\centering
	\includegraphics[width=\textwidth]{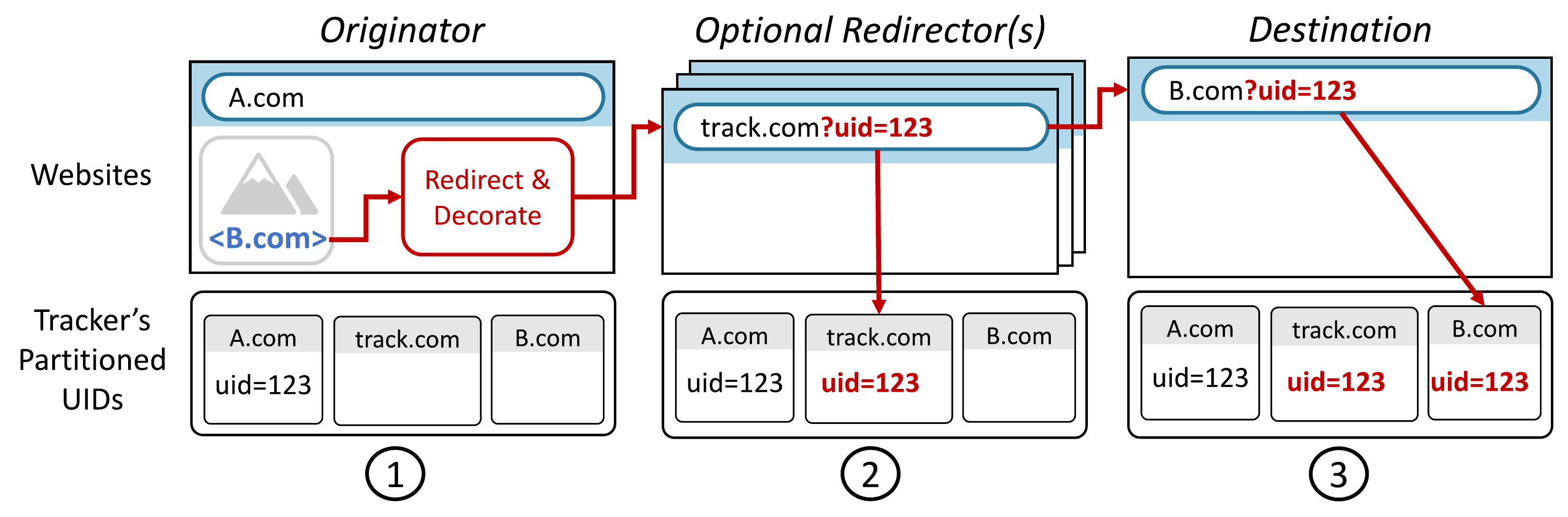}
	\caption{How \ids{} allows trackers to circumvent partitioned storage.}
	\label{fig:nav_tracking_explanation}
\end{figure*}

Figure~\ref{fig:nav_tracking_explanation} shows this process in
detail.  In \ids{}, the user is sent through a
\textbf{navigation path}.  This path begins at the \textbf{originator}
website, where the user clicks a link. When the link is clicked, the page 
itself or a tracker on the page decorates the URL by adding the originator's 
user identifier (UID) as a query parameter. The navigation path then passes 
through zero or more \textbf{redirectors}, which are invisible to the user but 
are permitted to store first party cookies. Each of these redirectors has the 
ability to store the UID from the query parameter as a cookie or local storage 
value under the redirector's domain. Finally, the user is sent to the 
\textbf{destination}, the website the link originally pointed to. The 
destination may also store the UID under its own domain. 
Thus, trackers using \ids{}{} regain the ability to share UIDs across websites 
with different domains, in defiance of the browser's partitioned 
storage protections.


%
\ids{} is related to, but more powerful than, two previously studied
tracking techniques: bounce tracking and cookie syncing. Bounce
tracking also modifies a user's navigation path by redirecting them through 
tracking sites that can store first-party cookies. Bounce tracking allows a 
tracker to record which originator and destination websites a user
has visited, but not to aggregate any information about a
user's behavior on those websites (the links the user clicks,
purchases the user makes, etc.), because no link decoration is used to insert 
UIDs into the navigation path. The 
tracker thus cannot link together the different UIDs it has assigned to a user 
across different websites. Both \ids{} and bounce tracking are part of a 
class of tracking techniques known as ``navigational tracking.''

Cookie syncing allows multiple third parties on a \emph{single} first-party
site to share UIDs with each other. However, if partitioned storage is in 
place, third parties cannot share information \emph{across} first-party 
websites using cookie syncing. 
When partitioned storage is in use, the storage available to trackers 
on the destination site is partitioned away from their storage on the 
originator. Thus, all trackers on the originator can share their UIDs with each 
other, and all trackers on the destination can do likewise, but trackers on the 
originator cannot share UIDs with trackers on the destination.

\section{Methodology}
\label{sec:methodology}

\begin{figure*}[t]
	\centering
	\includegraphics[width=\textwidth]{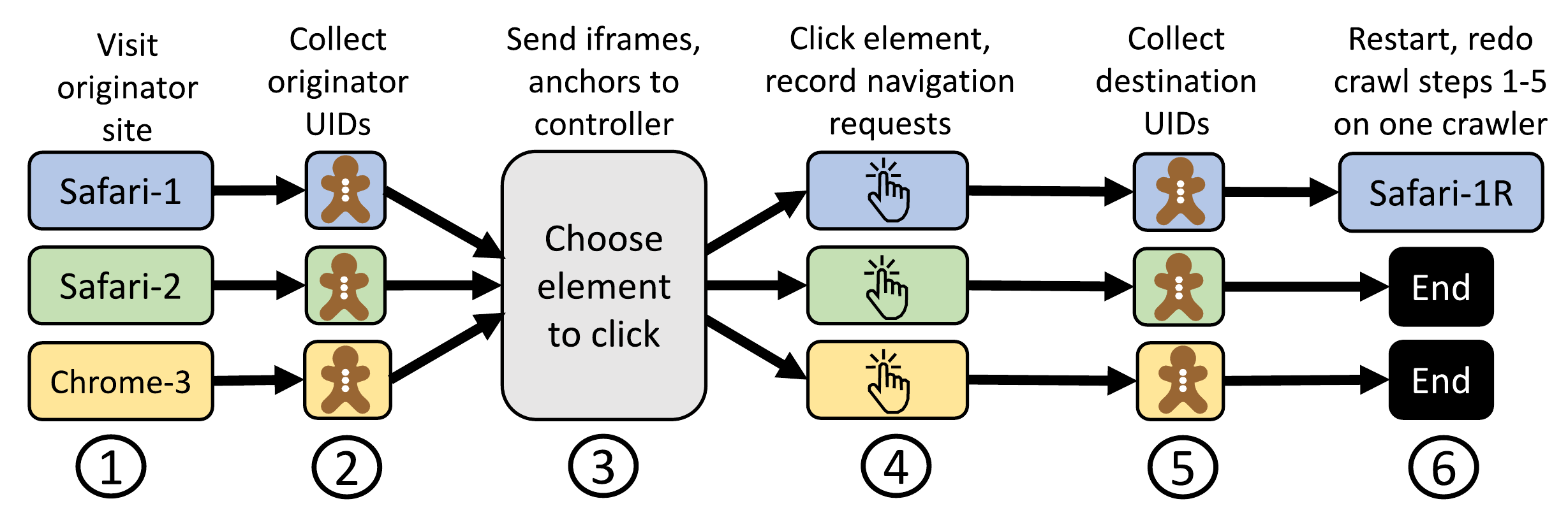}
	\caption{A single step of the ten-step random walk that CrumbCruncher 
		performs for each seeder domain.}
	\label{fig:crumbcruncher}
\end{figure*}

In this section, we describe CrumbCruncher, a web crawling system based on 
Puppeteer that measures the prevalence of \ids{} in the wild.
CrumbCruncher's goal is to collect as many potential cases of 
\ids{} as possible, and then  
distinguish the benign cases from true
\ids{} by determining which smuggled values are truly
UIDs. To collect potential \ids{}, CrumbCruncher employs multiple
synchronized crawlers that simulate a set of users, with an
additional, trailing crawler that simulates a user returning to each site.  
CrumbCruncher must then identify which potential UIDs are truly UIDs by 
comparing them across each crawl: values that vary across the set of different 
users but remain static for the repeat visitor are likely to be UIDs.

The canonical approach for identifying UIDs in prior work is to use only two 
crawlers to compare potential UID values across users. Unfortunately, these 
studies have been forced to discard a large number of potential UIDs from their 
analyses under two circumstances: first, when the potential UID only appeared 
on one crawler instead of both, and second, when the potential UID might be a 
session ID instead. 
Because we expect \ids{} to be rare and difficult to find in the wild, we 
require CrumbCruncher to discard as few UIDs as possible. 
CrumbCruncher achieves this goal in three ways. First, it 
distinguishes UIDs from session IDs more accurately than prior studies, which 
allows it to retain UIDs that would have been discarded by 
previous common strategies~\cite{koop2020depth, englehardt2015cookies, 
englehardt2016online, acar2014web}. Second, when potential \ids{} does not 
appear on 
all crawlers, CrumbCruncher applies programmatic and manual 
heuristics to identify UIDs, rather than discarding the cases entirely as prior 
work does~\cite{fouad2018missed, englehardt2015cookies, 
englehardt2016online,koop2020depth,acar2014web}. 
Finally, CrumbCruncher 
introduces a novel 
method for 
synchronizing web crawlers that click iframes, which allows it to collect data 
from the elements that are most likely to contain \ids{}. CrumbCruncher also 
uses three synchronized crawlers, rather than two, giving it multiple chances 
to observe each potential UID across two crawlers.
Each of these improvements allows CrumbCruncher to collect or retain more data 
than previous systems.


\subsection{Crawling the Web}

CrumbCruncher collects a sample of websites that contain \ids{} by performing 
ten-step random walks. Each random walk begins at a ``seeder domain'' taken 
from the Tranco list of the globally most-popular 10,000 domains~\cite{tranco}. 
Each of CrumbCruncher's multiple crawlers follows the same walk.

At each step of a walk, CrumbCruncher records all first-party cookies,
local storage values, and web requests on the originator page.  Next, it
chooses either a frame (\texttt{$<$iframe$>$}) or anchor
(\texttt{$<$a$>$}) element to click on, in an attempt to trigger
navigation.  CrumbCruncher selects iframes because we expect them to
contain advertisements which might use \ids{}. CrumbCruncher also
follows anchors because many webpages do not contain iframe elements.
Regardless of element type, CrumbCruncher prefers elements that
navigate to a URL with a different registered domain than the current
page.  For each click that triggers a navigation, CrumbCruncher's
browser extension collects all navigation web requests by implementing
a handler for the \texttt{chrome.webRequest.onBeforeRequest} event.
Upon arriving at the destination page, CrumbCruncher again records all
first-party cookies, local storage values, and web requests for ten
seconds.

CrumbCruncher repeats this navigation process, starting at each new page loaded 
by the click in the previous step, nine times. It then selects a new seeder 
domain to start the next random walk. 
CrumbCruncher retains browser state (including cookies 
and storage values) for the duration of each walk and discards it when beginning a new walk.
CrumbCruncher proceeds in this depth-first manner to
maximize the number of distinct pages  
visited, rather than maximizing the elements visited per page, because we 
expect that websites likely use \ids{} in either all elements 
that trigger navigation or none.

\subsection{Detecting potential \IDS{}}

CrumbCruncher's goal is to identify cases where a UID has 
been smuggled---i.e., passed across sites in defiance of browser 
protections---which requires differentiating UIDs from 
non-tracking tokens. We use the term ``token'' to 
refer to any potential UID found inside the value of a name-value pair, whether 
that pair is a first-party cookie, a local storage object, or a query 
parameter. CrumbCruncher builds on prior work that identifies UIDs by comparing 
the tokens that are passed by two different
users access a particular website~\cite{acar2014web, 
englehardt2015cookies, englehardt2016online, 
koop2020depth, fouad2018missed, urban2018unwanted}. However, instead of using 
two crawlers, CrumbCruncher uses four. 

Three of the four crawlers---nicknamed Safari-1, 
Safari-2, and Chrome-3---each simulate a different 
user on a Safari or Chrome browser. These three crawlers, which run in 
parallel, allow CrumbCruncher to discard tokens that are the same across users 
and are thus unlikely to be UIDs. We explain how CrumbCruncher spoofs browsers 
and impersonates 
different users in Sections~\ref{sec:impersonating_browsers} 
and~\ref{sec:impersonating_users}. 
The fourth crawler, Safari-1R, simulates the same user as Safari-1.
Safari-1R checks whether the same token is observed when a webpage is
accessed twice by the same user: specifically, Safari-1R repeats each
crawl step immediately after Safari-1 finishes it. Safari-1R allows
CrumbCruncher to discard tokens that differ when observed
repeatedly by the same user, and thus are probably session IDs, not
UIDs. More details on this process are given in
Section~\ref{sec:identifying_uids}.

\subsection{Synchronizing multiple crawlers}

One underlying assumption behind the multi-crawler methodology is that all
browsers are accessing the same version of a particular webpage: the four
crawlers must visit the same URL and click the same elements on each page.
This process is depicted in Figure~\ref{fig:crumbcruncher}. However, we find
that keeping the crawlers synchronized presents a significant challenge due
to the dynamic nature of the web. Determining which elements are the same on
different instances of the same webpage is not straightforward. Even when
accessed simultaneously, websites often load dynamic content: elements that
appear on one crawler's page might not appear on the others'. We also find
that even when elements are the same (e.g., iframes that load the same
content), they might not appear in the same locations or with the same size. 
Additionally, CrumbCruncher clicks iframe elements, which often do not have any 
attribute that identifies where a user will navigate when they click the iframe. 
Determining which iframe elements are equivalent across different instances of 
a webpage is more challenging than comparing anchor elements, which almost 
always have an easily comparable \texttt{$<$href$>$} attribute.

To mitigate this issue, CrumbCruncher uses a central controller (a local HTTP 
server) to 
choose the element that Safari-1, Safari-2 and Chrome-3 click in unison. 
Upon loading a page, each crawler sends a list of all 
anchor and iframe elements on that page to the central controller. These 
lists contain the elements' properties, location, bounding boxes, and x-paths. 
The controller compares the three lists to find elements that are the same 
across all three instances of the page. We consider elements to 
be the same if any of three heuristics are met:

\begin{enumerate}
	\item They are anchors and their \texttt{href} values are the same (not including 
	query parameters).
	\item They have the same property names (the values may differ) and similar 
	bounding boxes (the $y$-coordinate may differ, to allow for elements that 
	render at different heights on the page).
	\item They have the same property names and x-path.
\end{enumerate}

These heuristics are imperfect: they may incorrectly label elements as the same 
when they are not, or incorrectly discard elements. To mitigate these 
possibilities, CrumbCruncher compares the fully qualified domain name (FQDN) of 
the site each crawler has landed on at the end of every crawl step. If all 
three FQDNs are not the same, CrumbCruncher terminates the walk. 
We still include data from this last step in our analyses because this 
situation often 
occurs when CrumbCruncher has clicked on different advertisements that each 
exhibit a separate instance of \ids{}.

We evaluate the effectiveness of these heuristics and find 7.6\% of all
crawl steps fail because CrumbCruncher is unable to find an element that is
the same across all three synchronized crawls: this type of failure occurs at 
step \textcircled{3} in Figure~\ref{fig:crumbcruncher}. A further 1.8\% of 
crawl steps fail at step \textcircled{6} because the clicked elements were 
not actually the same, and led to different destination websites. 

%

\subsection{Impersonating different browsers}
\label{sec:impersonating_browsers}

All four crawlers use Chrome (version 95 or 92) 
because our chosen crawling framework, Puppeteer, is designed for that browser. 
However, CrumbCruncher impersonates Safari on three of our four crawlers by spoofing the 
\texttt{User-Agent} string.\footnote{We use the Safari 
\texttt{User-Agent} string \texttt{Mozilla/5.0 (Macintosh; Intel Mac OS X 10\_15\_7) 
AppleWebKit/605.1.15 (KHTML, like Gecko) Version/14.1.2 Safari/605.1.15.}} We 
chose to test Safari and Chrome because at the time of writing, Safari 
implemented partitioned storage by default, and Chrome did not. Our hypothesis 
was that trackers 
might use \ids{} on Safari to evade partitioned storage 
protections. Our Chrome-3 crawler was originally intended to test this 
hypothesis, but we were unable to use it for this purpose: \ids{} cases 
quite often appeared on only one crawler, regardless of whether that crawler 
was one of the three Safari crawlers or the Chrome crawler (see 
Section~\ref{sec:identifying_uids}). Differentiating cases where content that 
performed \ids{} was loaded dynamically from cases where \ids{} 
occurred deliberately on Safari and not Chrome proved to be impossible, so we 
simply use Chrome-3 as another distinct user to identify UIDs.

We note that while spoofing the \texttt{User-Agent} string does change the value of 
\texttt{window.navigator}, which is commonly used as a proxy for identifying 
the browser, it is not a foolproof method of impersonating a browser. 
Websites may use more sophisticated methods to identify a browser, such as 
comparing the codecs it supports~\cite{fp-crawlers}. However, 
relatively few websites go to such lengths: Vastel et al. crawled the 
Alexa 10K and found that only 93 websites appeared to use sophisticated 
fingerprinting techniques to identify the browser that was loading 
them~\cite{fp-crawlers}. We therefore consider the risk of sites misidentifying 
our browser to be small, given how few websites appear to use fingerprinting to 
identify browsers.

\subsection{Impersonating different users}
\label{sec:impersonating_users}

The Chrome browser differentiates users by storing profiles in a
folder called the ``user data directory''~\cite{user_data_dir}. To
simulate a new user at the start of each random walk, each crawler
starts with a new user data directory. This folder is modified from
the default in two ways: first, third-party cookies are disabled, and
second, a Chrome extension is installed that records web requests.

One potential limitation of our user simulation method is that
websites may generate UIDs using fingerprinting, i.e., by examining factors
like \texttt{User-Agent} string, supported fonts, hardware, and
more.\footnote{IP address is generally too variable to be used as an
input by fingerprinters~\cite{eckersley2010panopticlick}.}  Many of
these inputs are identical across all four crawlers since they are run
on one machine. If a tracker generated its UIDs using fingerprinting,
assigned the same UID across multiple crawlers, and then performed
\ids{}, CrumbCruncher would erroneously discard those cases. However,
we find that this rarely occurs by performing the following
experiment.

We observe that CrumbCruncher will not discard potential instances of 
\ids{} that only appear on a single crawler: only instances that 
appear on multiple crawlers and have identical UIDs will be 
discarded. If CrumbCruncher is erroneously 
discarding instances of \ids{}, we would expect to see 
very few cases that both occur on multiple crawlers \emph{and} 
originate on sites that perform fingerprinting.

To test this hypothesis, we separate cases of \ids{} into 
two groups: the cases that originate on sites that are known to employ
fingerprinting, and cases that originate on other sites. To determine which 
sites use fingerprinting, we use the list of fingerprinters found by Iqbal et 
al.~\cite{Iqbal2021FingerprintingTF}. We find that 13\% of 
\ids{} in our data originates on pages hosted by one of Iqbal et al.'s 
fingerprinters. We then divide both groups 
again, into the instances that occur on a single crawler and the instances 
that occur on multiple crawlers.  Next, we compare the proportion of 
single-crawler to multiple-crawler instances in the 
fingerprinting group to the 
non-fingerprinting group. In the fingerprinting group, 44\% of \ids{} cases 
occur on multiple crawlers, whereas in the 
non-fingerprinting group, 
52\% of cases occur on multiple crawlers.
While the
two-proportion Z test
suggests that this difference is
statistically significant---and, therefore, that
CrumbCruncher likely missed some cases of \ids{} due to fingerprinting---the difference is 
small.  The relative difference between populations suggests CrumbCruncher may 
have missed on the order of 13 
cases of \ids{} on sites that employ fingerprinting.
 
\subsection{Identifying Potential \IDS{}}

Once CrumbCruncher has finished collecting data, we search 
for potential UID tokens that were transferred across first-party 
contexts. We define ``different first-party contexts'' as 
the case when the site the token was originally found on has a 
different registered domain than any of the sites that eventually received the 
token, whether those sites are redirectors or the ultimate destination.

We extract potential UID tokens from cookies, local storage, and query 
parameters by 
recursively attempting to parse the value of each name-value pair\footnote{We 
do not look for tokens 
in the names of name-value pairs because Fouad et al. found that storing UIDs 
as names rather than values was a very uncommon 
practice~\cite{fouad2018missed}.} as JSON or URL-encoded 
values. For example, if a query 
parameter contains a JSON string that itself contains several URL-encoded 
tokens, we extract each URL-encoded token individually. 

We then discard all of the tokens that were not passed across at least one 
first-party context as a query parameter. For example, if the same token was 
found on both the originator site and the destination, but was 
not passed from the originator to the destination as a query parameter, we 
discard it. We find that the vast majority of these particular tokens are not 
used in \ids{}, but rather false positives that happen to appear on 
both websites. 

However, we keep tokens that only get passed across part of a 
navigation path. For example, if a token appears as a query parameter in the 
URL of a redirector, then gets passed to the destination, we keep it even if it 
did not appear on the original URL of that navigation path. Tokens are also not 
required to appear as cookies or local storage values: they can appear on the 
originator and destination as query parameters in third-party web requests. 

\subsection{Identifying UIDs}
\label{sec:identifying_uids}

After collecting all potential cases of \ids{}, we identify and discard all of 
the cases that transfer harmless values rather than UIDs. Examples of harmless 
values include timestamps, language specifiers, session IDs, and 
so on. While performing this analysis, we discovered that cases of potential 
\ids{} fell into two categories: we labeled these categories ``static'' and 
``dynamic.'' Static \ids{} occurs on elements that are always the same 
on every visit to the page. Consequently, cases of static \ids{} appear on all 
four crawlers. Dynamic \ids{} occurs on elements that load different content on 
different page visits. Cases of dynamic \ids{} appear on fewer than all four 
crawlers, despite our efforts to keep the crawlers synchronized. 
Identifying UIDs in static \ids{} is simpler than in dynamic \ids{}: we 
describe our procedure in the static case first.

\subsubsection{Identifying UIDs in the static case}
\label{sec:identifying_uids_static}

To track an individual user, a UID must be the same across all website visits 
by the same user and different across visits to the same website by different 
users. Consequently, we discard any token that is the same across our crawlers 
that simulate different users, since these cannot be UIDs. However, it is also 
necessary to discard tokens that differ across a single user, since these 
tokens are likely to be session IDs that are not used for user tracking.
Prior work discarded session IDs by discarding all 
tokens whose lifetime was less than a specific time, such as 90 
days~\cite{koop2020depth, englehardt2015cookies, englehardt2016online} or a 
month~\cite{acar2014web}. CrumbCruncher improves on prior work by 
comparing potential session IDs across Safari-1 and Safari-1R, which simulate 
the same user visiting the same website twice, and discarding the tokens that 
differ across these crawlers.
%
A sampling of data collected from one of our crawler machines indicates that
16\% of the UIDs we identify have
a lifetime of less than 90 days, and 9\% have a lifetime shorter than a month. 
These UIDs would have been missed by prior work that uses lifetime to determine 
whether a token is a session ID.

\subsubsection{Identifying UIDs in the dynamic case}
\begin{table}
	\begin{tabular}{lr}
		\toprule
		User Profiles & \# Tokens \\
		\midrule
		2 identical plus 1 or more different profiles & 325 \\
		2 or more different profiles only & 171 \\
		2 identical profiles only & 20 \\
		1 profile only & 445 \\
		\bottomrule
	\end{tabular}
	\caption{Crawler combinations where UIDs appeared.}
	\label{tab:certainty_levels}
\end{table}

Unfortunately, we found that the majority of potential \ids{} 
instances were dynamic and thus did not occur on all four crawlers: in fact, 
many instances occurred on only a single crawler. For example, we encountered 
many cases where each crawler loaded the same 
originator website and clicked the same iframe element, but the iframes 
contained different advertisements, so each advertisement presented a different 
navigation path and arrived at a different 
destination. We classify tokens that appear on fewer than four 
crawlers in the following manner:

\begin{enumerate}
	\item If a token is present in any two crawls with 
	different user profiles, and its value is the \emph{same} 
	across those crawls, we discard it.
	\item If a token's name is present in Safari-1 and 
	Safari-1R, which have the same user profile, and its value 
	\emph{differs}, we discard it.
\end{enumerate}

\noindent We are left with two classes of tokens: tokens that are present 
in only a single crawl, and tokens that only appear in crawls 
with different profiles (and have different values across each 
crawler). For these tokens, we employ both the programmatic 
heuristics used by previous work and manual sorting. 

We base our programmatic heuristics on those 
of prior studies~\cite{englehardt2015cookies, englehardt2016online, 
koop2020depth}. We remove tokens that appear to be dates or 
timestamps, tokens that appear to be URLs, and tokens that are 
less than seven characters long. We do not impose any 
restrictions based on cookie expirations. However, even after 
we applied these filters, manual inspection of the remaining 
tokens revealed a high number of obvious false positives. These 
included natural language strings separated by delimiters 
(such as ``Dental\_internal\_whitepaper\_topic,'' 
``share\_button''), concatenated 
words with no delimiter (``sweetmagnolias,'' ``trustpilot''), 
semi-abbreviated words (``navimail''), acronyms (``en-US''), 
and more. Filtering most of these 
out programmatically presented a significant challenge.  

We therefore concluded that programmatic heuristics would be 
insufficient to distinguish UIDs from other tokens, 
and resorted to removing obvious false positives by hand. Our 
final, conservative strategy is to remove tokens that 
are composed of any combination of  natural language words, 
coordinates, domains, or obvious acronyms like 
``en-US.''  
Table~\ref{tab:certainty_levels} shows how many of the final set of
UIDs were present 
on various combinations of crawlers. 

In the end, we manually removed 577 out of 1,581 tokens because our 
programmatic filters failed to 
recognize them as non-UIDs. This number is significantly higher 
than we expected and underscores the value of attempting to 
observe UIDs across as many crawlers as possible. 
Care should therefore be taken when comparing 
our results to prior work that did not attempt to manually 
remove false positives.

\subsection{Implementation}

We implemented CrumbCruncher using both Puppeteer, to automate site visits 
and record cookies and local storage, and a custom Chrome extension, to record 
web requests. We use Puppeteer in ``headful'' mode, using the monitor emulator 
XVFB~\cite{xvfb}, 
to reduce the chance that CrumbCruncher will be identified as a bot. While 
Puppeteer is capable of recording most web 
requests, it cannot guarantee that it can attach request handlers before any 
requests on a page have been sent~\cite{puppeteer_bug_3667,puppeteer_bug_2669}. 
We found during initial testing
that this led to a significant number of missed requests; hence,
CrumbCruncher records requests using a browser extension
instead. CrumbCruncher runs on twelve Amazon EC2 \texttt{t2.large}
instances. Each EC2 instance has a different set of 834 seeder
domains. The full crawl of 10,000 seeder domains takes approximately
three days to complete.

\begin{table}[t]
  \centering
  \begin{tabular}{lr}
    \toprule
    Unique URL Paths & 10,814\\
    \hspace*{0.2in} Unique URL Paths w/ UID Smuggling & 850 \\
    \hspace*{0.2in} Unique Domain Paths w/ UID smuggling & 321 \\
    Unique Redirectors & 214 \\
    \hspace*{0.2in} Dedicated Smugglers & 27\\
    \hspace*{0.2in} Multi-Purpose Smugglers & 187\\
    Unique Originators & 265 \\
    Unique Destinations & 224 \\
    \bottomrule
  \end{tabular}
  \caption{Summary of the navigation paths and their 
  participants measured by CrumbCruncher.}
  \label{tab:summary}
\end{table}

\section{Results}
\label{sec:results}

We consider two forms of navigation paths in our evaluation. ``URL 
paths'' consist of the full URLs of the originator, any redirectors, and the
destination (e.g., 
\texttt{a.com/x/y?UID=0}~$\rightarrow$ \texttt{b.com/x/y?UID=0}).  Domain paths 
consist only of the domains at each step of the path (e.g., 
\texttt{a.com}~$\rightarrow$ \texttt{b.com}).

In total, we observed 10,814 unique URL paths in the data set we gathered using CrumbCruncher.
We consider unique
URL paths, rather than total URL paths including duplicate 
paths, because this metric gives 
a better estimate of how many websites participate in \ids{}.

Using our method for identifying UIDs, we
found \ids{} on 8.11\% of the unique URL paths 
taken by CrumbCruncher. It is interesting that such a 
non-trivial percentage of advertisers
have implemented \ids{}, especially given that Chrome---the most
widely used web browser---still permits tracking with 
third-party cookies by default. 
We speculate that the affiliate advertising 
market may be driving the adoption of \ids{}: affiliate programs 
have reportedly been failing to attribute conversions because of browsers' 
third party cookie blocking~\cite{tune_server_side}, and link decoration allows 
conversions to be attributed correctly.

In the rest of this section we examine the \ids{} we
discovered in detail to understand who is implementing it, how they implement
it, and why they implement it.


\begin{table}[t]
	\begin{tabular}{lrr}
		\toprule
		Redirector & Count & \% Domain Paths \\
		\midrule
		adclick.g.doubleclick.net & 36 & 11.2 \\
		googleads.g.doubleclick.net & 20 & 6.2 \\
		advance.lexis.com* & 10 & 3.1 \\
		d.agkn.com & 9 & 2.8 \\
		btds.zog.link & 9 & 2.8 \\
		ad.doubleclick.net & 8 & 2.5 \\
		gm.demdex.net & 8 & 2.5 \\
		www.kinopoisk.ru* & 7 & 2.2 \\
		secure.jbs.elsevierhealth.com & 6 & 1.9 \\
		t.myvisualiq.net & 6 & 1.9 \\
		11173410.searchiqnet.com & 6 & 1.9 \\
		optout.hearstmags.com* & 6 & 1.9 \\
		signin.lexisnexis.com* & 6 & 1.9 \\
		trc.taboola.com & 5 & 1.6 \\
		l.instagram.com* & 5 & 1.6 \\
		ads.adfox.ru* & 5 & 1.6 \\
		www.facebook.com* & 5 & 1.6 \\
		reseau.umontreal.ca* & 5 & 1.6 \\
		l.facebook.com & 4 & 1.2 \\
		rtb-use.mfadsrvr.com & 4 & 1.2 \\
		www.campaignmonitor.com* & 4 & 1.2 \\
		6102.xg4ken.com* & 4 & 1.2 \\
		swallowcrockerybless.com* & 4 & 1.2 \\
		montreal.imodules.com* & 4 & 1.2 \\
		www.getfeedback.com* & 4 & 1.2 \\
		kuwosm.world.tmall.com* & 4 & 1.2 \\
		www.awin1.com & 3 & 0.9 \\
		www.zenaps.com & 3 & 0.9 \\
		pr.ybp.yahoo.com & 3 & 0.9 \\
		go.dgdp.net & 3 & 0.9 \\
		\bottomrule
	\end{tabular}
        \vspace{0.75\abovecaptionskip}
	\caption{The most common redirectors observed in 
		unique domain paths. *Multi-purpose smuggler}
	\label{tab:redirectors}
	\vspace*{-0.15in}
\end{table}

\subsection{Redirectors}
\label{sec:trackers_vs_potential}

\begin{figure*}[t]
	\centering
	\includegraphics[width=\textwidth]{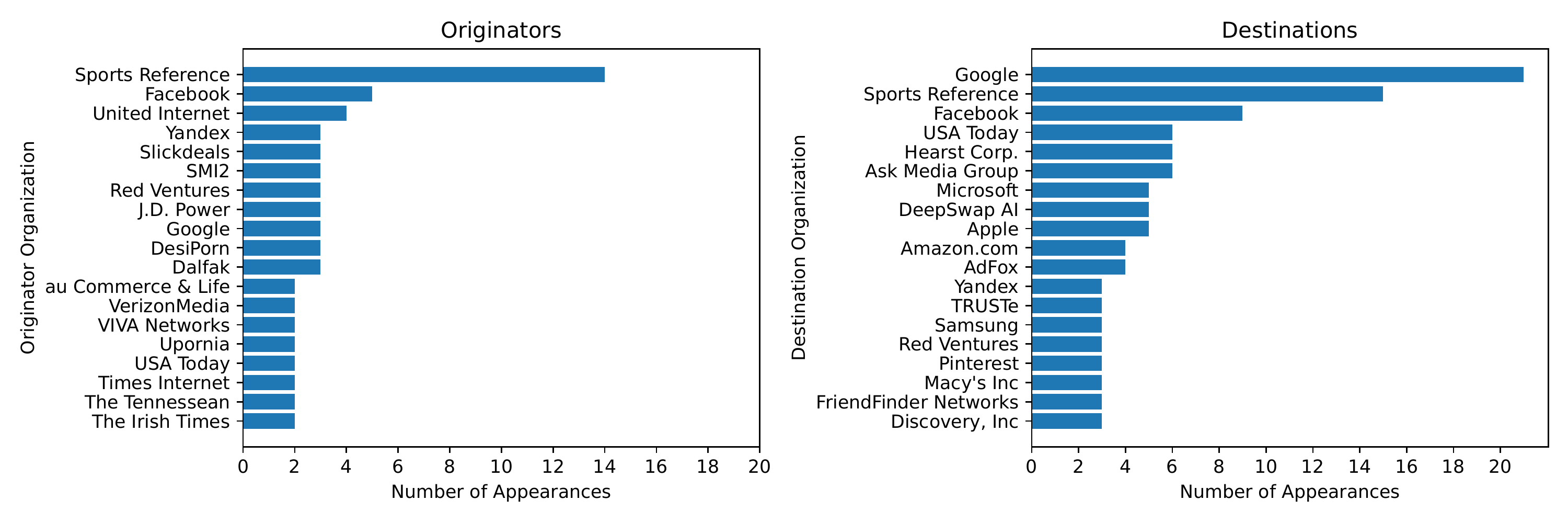}
	\caption{Most common entities involved in \ids{} as originators or 
	destinations.}
	\label{fig:culprits}
\end{figure*}

We start by identifying the trackers involved as redirectors in the
navigation paths that include \ids{}.
We use unique domain paths instead of URL paths for this analysis, because this 
metric better captures how widely a redirector is spread across the web, 
without 
over-counting repeated instances of \ids{} by the same entity.

We classify redirectors into two groups: 
``\dsmugglers{}'' and ``multi-purpose smugglers.'' We use a 
conservative heuristic to identify
\dsmugglers{} that appear to have no purpose in the navigation path 
besides \ids{}.\footnote{For example, site-specific redirection services (e.g., 
Twitter's \texttt{t.co}) are not considered \dsmugglers{} using this 
classification.} 
We consider a redirector a \dsmuggler{} if it meets three 
requirements:
\begin{itemize}
	\item The redirector appears in navigation paths whose 
	originators have multiple different registered domains,
	\item The redirector appears in navigation paths that end in destinations 
	with multiple registered domain names,
	\item The redirector's FQDN is never observed as an 
	originator or destination.
\end{itemize}

We separate out \dsmugglers{} because we are confident that these domains have 
no purpose besides \ids{}, and their sole intent is therefore likely to be 
enabling trackers to aggregate users' information 
across websites. We also predicted that \dsmugglers{} 
might be particularly underrepresented in filter lists that block trackers, 
because \ids{} is such a recent technique. 
Indeed, when we compared the \dsmugglers{} that we 
found to the Disconnect list of trackers~\cite{disconnect_trackers},
41\% of them (11 out of 27) were not yet present in the list. 

However, our heuristic is conservative. The less often CrumbCruncher 
sees a redirector, the less likely it is to observe multiple 
originators and destinations for that redirector, in which case
the redirector would not be classified as a \dsmuggler{}. 
Consequently, some \dsmugglers{} might appear in the ``multi-purpose smugglers'' 
category.



Table~\ref{tab:redirectors} shows the most commonly-occurring
redirectors in the navigation paths we measured.  From this list, 16
of the 30 most common redirectors are \dsmugglers{} and 14 are multi-purpose 
(the multi-purpose smugglers are marked with an asterisk). Of the 16 
\dsmugglers{}, 14
are owned by advertisers, while the other two
(\texttt{btds.zog.link} and \texttt{secure.jbs.elsevierhealth.com}) have 
unclear owners or purposes. The most
commonly used \dsmuggler{} is DoubleClick, which appears in more than
20\% of all cases of \ids{}.

The multi-purpose trackers appear to fill a variety of rolls: while all of 
them perform \ids{}, some have a separate purpose as well. Some redirect 
to sign-in pages (e.g., \texttt{signin.lexisnexis.com}), host user-facing 
websites (e.g., \texttt{www.facebook.com}), upgrade or 
downgrade HTTP/HTTPS connections (e.g., \texttt{kuwosm.world.tmall.com}),
or specify the English-language version of a 
site by appending ``/en/'' (e.g., \texttt{www.getfeedback.com}). Some 
multi-purpose smugglers are owned by advertising companies, just as the 
\dsmugglers{} are. Two redirectors, 
\texttt{swallowcrockerybless.com} and \texttt{d.agkn.com}, appear to be 
associated with adware or other potentially unwanted activity. 

\subsection{Originators and Destinations}

Next, we identify the organizations that acted as originators or
destinations during \ids{}. We began with the Disconnect
entity list~\cite{disconnect_entity_list}, which recorded an owning 
organization for 45 out of the 436 unique registered domains of the originators 
and destinations. We then identified the owners of a further 235 registered 
domains manually (all of the domains that appeared multiple times, plus as many 
of the long tail as we could). An entity is counted once per unique
domain path: if multiple domains owned by a single organization appear more 
than once in a domain path, the owning organization is only counted once for 
that path.

Figure~\ref{fig:culprits} shows the originators and destinations observed most 
frequently in our measurements. We note that many originators might be expected 
to publish affiliate advertisements, such as sports websites, news 
organizations, and adult websites, while many destinations might have affiliate 
advertising programs, such as retailers or technology companies. While we 
cannot guarantee that these entities participate in \ids{} as part of affiliate 
advertising campaigns, many of these organizations support that hypothesis.

Figure~\ref{fig:culprits} also illustrates one particular case of \ids{} 
between unexpected organizations. One of the most 
common cases of \ids{} in our measurements was a navigation 
path that led from the originator \texttt{instagram.com}, owned by Facebook, to 
the Google Play Store.
This path existed 
because the button on \texttt{instagram.com} advertising the Instagram 
mobile app always appended \texttt{instagram.com}'s UID 
cookie to the navigation request for \texttt{play.google.com}. We were 
surprised to see that two large advertising companies, that might be expected 
to be competitors, were apparently sharing UIDs with each other. 

Figure~\ref{fig:culprits} also contains an example of \ids{} that was not 
initiated by an advertiser, but rather used to synchronize information between 
multiple domains owned by a single company. The most 
common originator in Figure~\ref{fig:culprits} is Sports Reference, an
organization that maintains several websites with statistics for popular
American sports. This company owns several sports-themed domains whose websites 
link frequently to each other, such as stathead.com, hockey-reference.com, 
baseball-reference.com, and others~\cite{sports_ref}. CrumbCruncher spent 
several random walks in this ecosystem of websites. We hypothesize that rather 
than using \ids{} for advertising, Sports Reference uses it to share 
information between its own affiliated sites.

\subsubsection{Content categories}
\begin{figure}[t]
	\centering
	\includegraphics[width=3in]{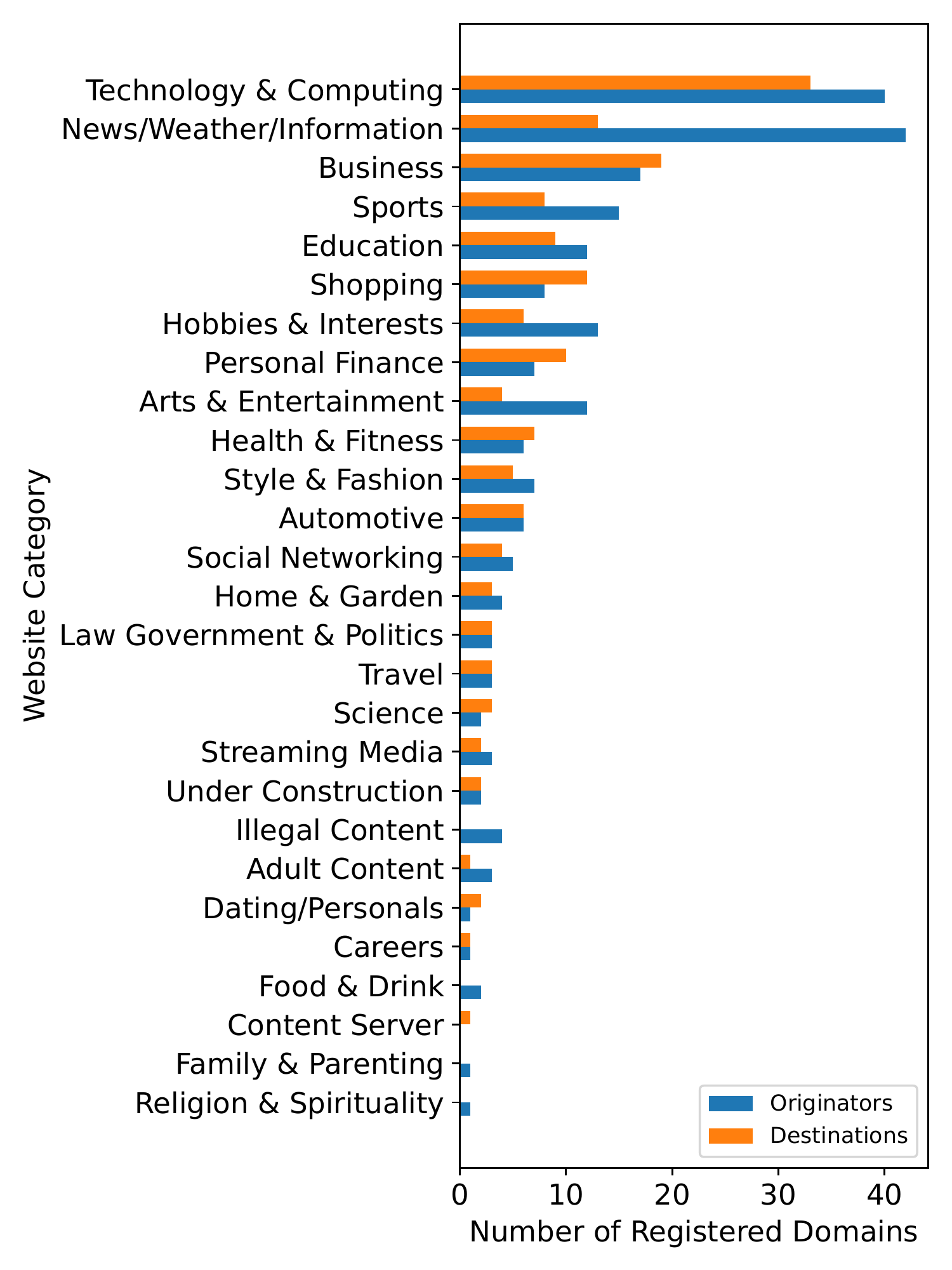}
	\caption{ Categories of websites that participate in 
		\ids{} as originators or destinations.}
	\label{fig:categories}
\end{figure}

We further break down the originators and destinations by
categorizing them by the topic of their site content. We use the 
categorization defined by the IAB 
Tech Lab Content Taxonomy~\cite{iab_dataset} as provided by
Webshrinker~\cite{webshrinker}, whose data set
contains 404 domain categories~\cite{iab_categories}.
Out of 339 unique registered domains, 307 had a useful
category and 32 were categorized as unknown.

Figure~\ref{fig:categories} shows the most common categories of
websites that participate in \ids{} in our dataset. The
counts of websites per category reflect the number of unique registered domains 
in that category, so that each registered domain is represented only once even 
if CrumbCruncher encountered it multiple times. For example, even though
Facebook's domains are common originators as seen in Figure~\ref{fig:culprits},
they only appear twice as originators in Figure~\ref{fig:categories}: once for 
\texttt{facebook.com} and once for \texttt{instagram.com}, both in
the ``Social Networking'' category.

Notably, ``News/Weather/Information'' is the most common category for
originators, and the second most common category overall. This result
is consistent with previous studies that found news websites to have
an above-average amount of more traditional tracking mechanisms, such
as fingerprinting and tracking
pixels~\cite{englehardt2016online,Iqbal2021FingerprintingTF}.
Our impression, based on manual inspection of a few of these originators, is 
that news websites have an above-average number of advertisements in
iframes that perform \ids{} when clicked.


\subsubsection{Third parties}

\begin{figure}[t]
	\centering
	\includegraphics[width=\columnwidth]{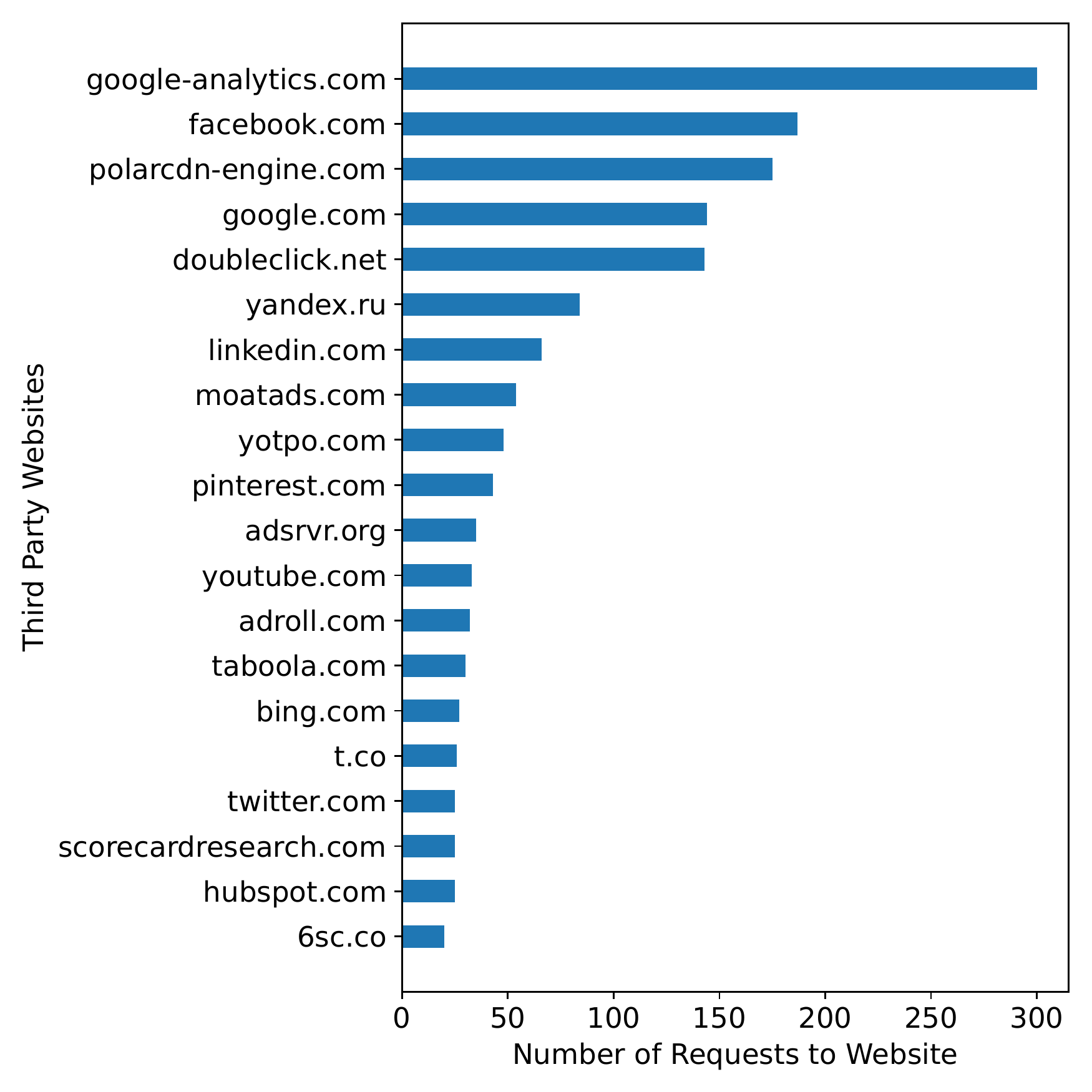}
	\caption{Most common domains of third party web requests sent from the 
	destination site.}
	\label{fig:dst_other}
\end{figure}

After a UID has been transferred through the entire 
navigation path, it may not have finished its journey: third 
parties on the destination site may also send the UID back to 
their own servers. Figure~\ref{fig:dst_other} shows the 
20 most common registered domains of the targets of web requests 
sent from destination sites that included UIDs. 

The third-party trackers listed in this 
figure include trackers that did not appear to use \ids{}. We note that 
many requests to third party trackers passed the UID only because the request 
included the entire URL of the destination site, suggesting that the UID may 
have been ``leaked'' to these entities accidentally. This unintended 
consequence of \ids{} may present a further privacy harm, in that trackers that 
do not participate in \ids{} are nevertheless gaining access to UIDs that they 
would otherwise be unable to observe.

\subsection{Navigation Paths}

\begin{figure}[t]
	\centering
	\includegraphics[width=3in]{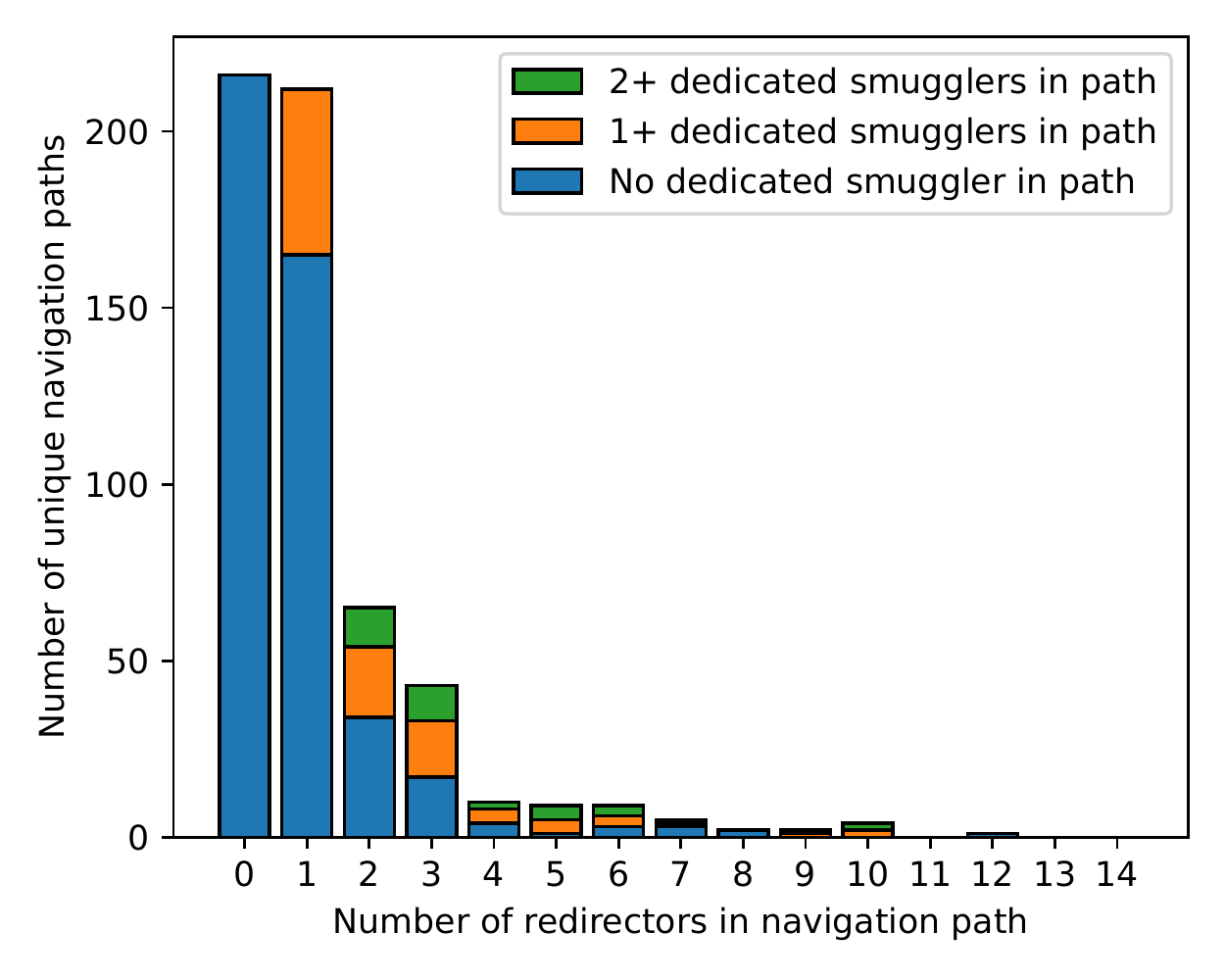}
	\caption{Distribution of the types of redirectors in URL paths.}
	\label{fig:chain_lengths}
\end{figure}

\begin{figure}[t]
	\centering
	\includegraphics[width=3.5in]{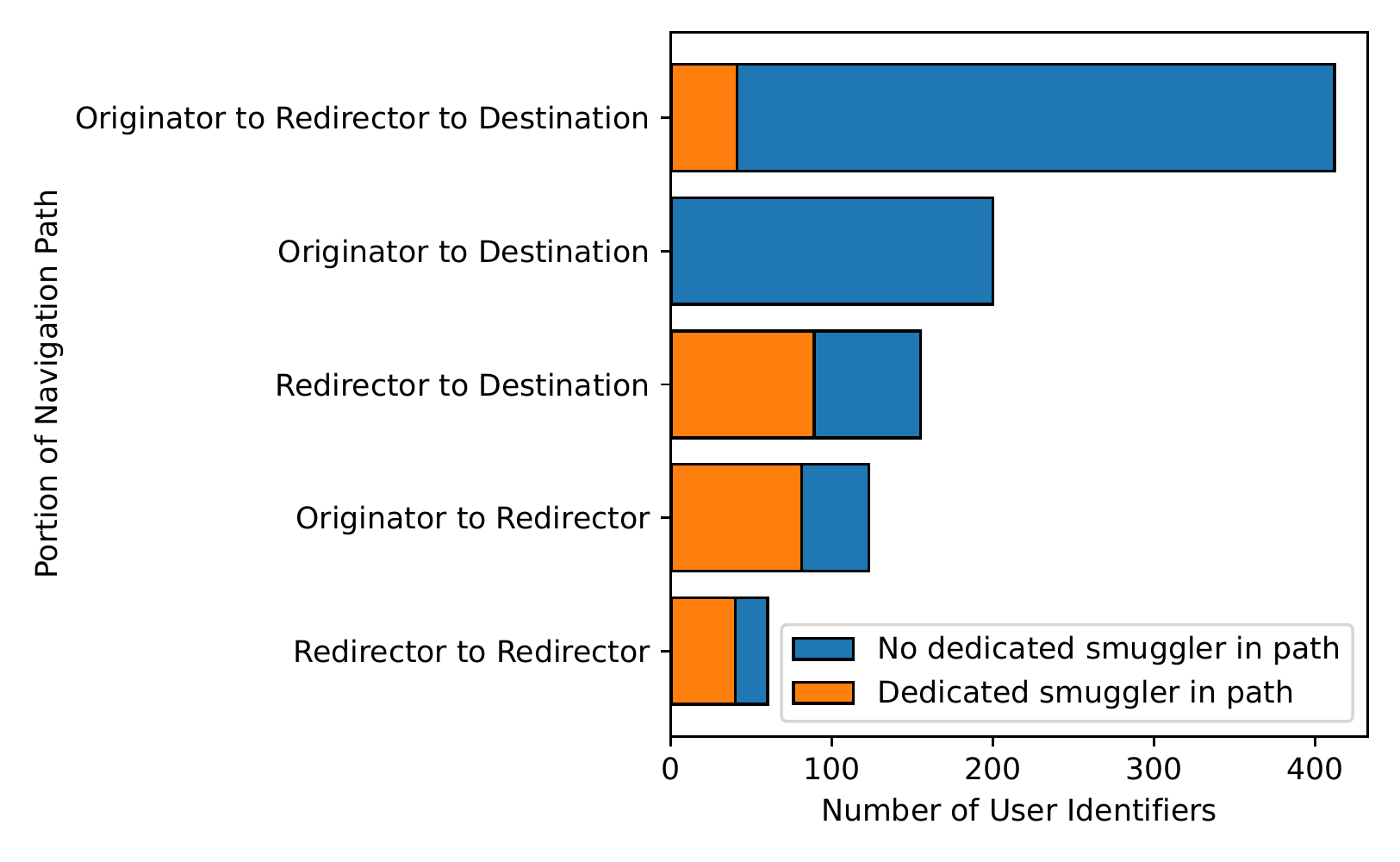}
	\caption{Counts of UIDs that traversed each portion of a URL
	path.}
	\label{fig:token_transfer_paths}
\end{figure}

In this section, we examine the characteristics of navigation paths used for 
\ids{}, including the features that differentiate them from benign navigation 
paths. 

Figure~\ref{fig:chain_lengths} shows the number of 
redirectors in the middle of each URL path that was used for 
\ids{}. The first bar, with zero redirectors, 
shows the cases where a UID was transferred directly from the 
originator to the destination without passing through any 
redirectors in between. 

The results show that the higher the number of redirectors 
in a path, the greater the proportion of paths that contain \dsmugglers{}, and 
the greater the number of \dsmugglers{} in each path. We conclude that shorter 
navigation paths are more likely to have a benign purpose, whereas longer 
navigation paths are more likely to be used for \ids{}.

Long navigation paths give multiple trackers the ability to share UIDs 
with each other. For example,
one navigation path started at a coupon-collecting website
(\texttt{couponfollow.com}), passed through a partner site 
owned by the same entity, then passed through four different trackers
before arriving at the final destination (a retailer). Each of these
trackers had the ability to record information about the ad the user
had clicked and their apparent interest in the retailer's products. 

Long navigation paths can also allow a single tracker to coordinate multiple 
domains that it controls. If those domains are connected to 
separate infrastructure (as might be the case if one advertising company 
acquired another and inherited the acquisition's infrastructure), the company 
might wish to synchronize the UIDs stored as first party cookies by 
redirectors. For example, the most common pair of redirectors we 
observed (where the first domain in the pair immediately redirects to the 
second domain) is \texttt{awin1.com} $\rightarrow$ \texttt{zenaps.com}. Both 
domains are owned by the advertiser AWIN.

UIDs do not always begin at the originator and 
pass through each redirector before arriving at the destination: they may 
appear at any step of the path and cease their journey at any number of hops 
further along. Each navigation path can also contain multiple UIDs. 
Figure~\ref{fig:token_transfer_paths} shows how many UIDs 
traverse each portion of the navigation path. We divide the UIDs that 
traverse each partial path into two groups: the UIDs that passed through a 
\dsmuggler{}, and the UIDs that passed through either multi-purpose smugglers 
only or no redirectors at all. For example, the second 
bar, ``Originator to Destination,'' shows the number of UIDs that were passed 
through navigation paths that did not include any redirectors.

We observe that the majority of UIDs are transferred across the entire path 
from the originator, through any redirectors if they exist, to the destination. 
A tracker might wish to do this when it is reasonably confident that the 
destination will include one of its scripts, which is capable of storing the 
UID under the destination's domain. If a tracker is present on the originator 
and capable of initiating \ids{}, but is 
not confident that the destination will contain one of its scripts, it might 
choose to transfer the UID through only part of the navigation path. These 
``partial transfer'' cases involve a higher proportion of \dsmugglers{}, which 
is further confirmation that the redirectors we label ``dedicated'' have no 
other purpose in the navigation path than \ids{}. We hypothesize that 
trackers who only send a UID through a part of the navigation path might be 
less widely used, since they are apparently not confident that the destination 
will contain one of their scripts.

\section{Limitations}
\label{sec:limitations}

CrumbCruncher has several limitations. First, we only look 
for UIDs that are transferred in the query parameters of 
URLs, and not by other methods. For example, trackers 
reportedly sometimes decorate the link in the document.referrer header with the 
UID, instead of the link to the destination 
page~\cite{webkit_itp_2_3}. Our initial reasoning was that 
there are a wide variety of ways to transfer UIDs, so we could simply check 
once a crawl was complete for UIDs that had mysteriously appeared on different 
websites without being passed through a URL. In practice, 
this turned out to be difficult: dynamic instances of \ids{} had to be detected 
using heuristics, which gave large numbers of false positives when used without 
the additional information provided by multiple crawlers. It turned 
out that when the same value appeared on two different 
websites, the most common reason was that the value was 
not a UID and had simply happened to be generated on both 
sites. To reduce our false positive rate and therefore the number of 
identifiers we had to remove by hand, we chose to 
consider only values that we had observed get transferred 
across at least two first party contexts.

Second, if a website uses browser fingerprinting to 
generate UIDs, our methodology may not fool the site into 
believing that our crawlers represent different users. We have calculated that 
the effect of browser fingerprinting on our results is very small: please see 
Section~\ref{sec:impersonating_users} for details.

\section{Countermeasures}
\label{sec:countermeasures}

\subsection{Existing Mitigations}

Defending against \ids{} is not 
straightforward. 
Given the difficulty of designing defenses that do not 
degrade user 
experience, most defenders (whether browsers or browser extensions) have so far 
opted for either heuristic-based or blocklist-based 
approaches.
For example, Safari uses heuristics: the browser will 
delete cookies and website 
data set by a redirector unless the
user also interacts with the redirector as a first-party
website~\cite{safari_cross_site_tracking}. Safari 
labels an originator as performing 
\ids{} if 1) it automatically redirects the user to another site,
and 2) it did not receive a user 
activation~\cite{w3c_nav_tracking}. Safari also 
classifies a site as a UID smuggler if it 
participates in a navigation path that contains another 
known UID smuggler. 

In contrast, Firefox defends against UID smugglers using the Disconnect
Tracker Protection blocklist~\cite{disconnect_trackers, 
mozilla_protection}. Firefox
clears all storage from sites on the Disconnect 
tracking list after 24 hours, unless the user has loaded
the site as a first party in the previous 45 
days~\cite{w3c_nav_tracking}. Unfortunately, we found that many UID smugglers 
are not yet present on the Disconnect list.

The Brave browser has multiple approaches for preventing \ids{}. First, if the 
browser is navigating to a link with a query parameter for another
destination URL, Brave will simply redirect to the URL in the query
parameter~\cite{brave_debouncing}. If the browser cannot detect the final 
destination of the navigation, it allows the navigation to proceed, but inserts 
an interstitial that warns users they will be tracked if they continue. Brave 
also maintains a list of \ids{} URLs created from crowd-sourced and 
open-source information, as well as a blocklist of query parameter names that 
are commonly used for \ids{}. Finally, 
Brave clears the storage areas associated with any sites it classifies as 
UID smugglers as soon as the user closes the tab that loaded them.

While Chrome is in the process of deprecating third-party
cookies~\cite{chrome_privacy_sandbox}, it does not appear to implement any
features to defend against \ids{} yet.

Some browser extensions have begun to implement protections against 
\ids{} as well. For example, Privacy 
Badger~\cite{privacy_badger} -- a browser extension by
the Electronic Frontier Foundation that blocks 
cross-site tracking -- 
identifies when a tracker inserts a redirector into a navigation path, and 
extracts the destination link from the query parameter in the
redirector's URL~\cite{privacy_badger_fb_tracking}. Another extension, uBlock 
Origin, implements an interstitial-based approach similar to 
Brave's~\cite{newman_incredibly_2021}.

\subsection{Proposed Mitigations}

CrumbCruncher's data can help augment the blocklists 
used by privacy tools and browsers to defend against 
\ids{}. We provide two contributions: 
first, we commit to 
publishing our list of token names and trackers in a 
publicly available GitHub repository (although we will 
refrain from recording the link here at this time to 
preserve the double-blind nature of the review 
process). This repository will 
contain the list of query parameter names that 
were used to transfer UIDs across 
websites, as well as the list of entities 
that participate in \ids{} as 
redirectors. Our second contribution is the code for 
CrumbCruncher itself, which 
can be run as an almost entirely automated pipeline in 
order to continuously update blocklists of navigational 
trackers. A major challenge of blocklist-based defenses lies in 
keeping those blocklists up to date: CrumbCruncher can help do that 
with much less human intervention than systems that rely on user reports of 
\ids{}. We will publish the code for CrumbCruncher along with the list of token 
names and UID smugglers that we discovered in this study.

\section{Related Work}
\label{sec:related}


The work that is most closely related to our own is Koop et al.'s 
study of bounce tracking~\cite{koop2020depth}. Bounce tracking is similar to 
\ids{} in that users' navigation paths are modified to insert redirectors that 
can store values as first parties, but differs in that no UIDs are transferred 
across contexts. Koop et al. study bounce tracking only, and do not attempt to 
measure whether UIDs are transferred across contexts. CrumbCruncher also clicks 
both iframes and anchors, whereas Koop et al.'s crawler clicks only 
anchors. This allowed CrumbCruncher to detect navigational tracking 
used by advertisements in iframes. 

To verify that CrumbCruncher crawled a reasonable sample of the Web and 
successfully detected modified navigation paths, we measured the instances of 
bounce tracking that CrumbCruncher observed while it searched for \ids{}, and 
compared our findings to the instances found by Koop et al. We found that 
bounce tracking that did not also involve \ids{} was present on 2.7\% of the 
navigation paths we studied (\ids{} was present on 8.1\%). Because Koop et al. 
did not attempt to measure whether UIDs were transferred across contexts, their 
study labeled all \ids{} that involved one or more redirectors as bounce 
tracking. Koop et al. found that ``11.6\% of the websites in the Alexa top 
50,000 had at least one link leading to one of the top 100 [most common] 
redirectors.'' This finding seems consistent with our measurement that either 
\ids{} or bounce tracking is present on a total of 10.8\% (8.1\% \ids{} and 
2.7\% bounce tracking) of the unique navigation paths we followed.

\subsection{Prior work on differentiating UIDs}
Multiple groups have attempted to differentiate between identifiers 
that are capable of tracking users (UIDs) and identifiers that are 
not. To be a UID, a value must 
differs across different users, remain the same for the same user (i.e., it 
must not be a session ID), and contain sufficient entropy. Techniques for 
making these three determinations vary.

Prior work, which focused on cookies that might be UIDs, determined whether a 
cookie varied across users by
directly or indirectly simulating different users across different 
crawls. Some work used two crawlers that visited the same sites 
simultaneously~\cite{englehardt2016online, englehardt2015cookies, 
fouad2018missed}, while others simulated multiple users sequentially 
using a single crawler ~\cite{koop2020depth} or multiple 
crawlers~\cite{urban2018unwanted}. Simulating multiple users 
sequentially enables a crawler to simulate more different users, because 
keeping multiple crawlers synchronized becomes more difficult as the number of 
crawlers increases, and a single crawler can evade this problem entirely. The 
disadvantage of sequential user simulation in prior work is that the crawlers 
did not guarantee that they visited each website more than once and thus 
observed each cookie more than once. Consequently, some of the cookies measured 
by the single sequential crawlers could not be compared across multiple users. 
In contrast, CrumbCruncher makes a concerted effort to visit every website in 
each crawl with four crawlers that represent three different users, which 
maximizes the chance that we can compare cookies and local storage values 
across users. 

Determining whether a token is a UID also requires discarding session 
IDs. Most past studies labeled cookies as session IDs if their  
lifetime was less than a specific time, such as 90 
days~\cite{koop2020depth, englehardt2015cookies, 
englehardt2016online} or a month~\cite{acar2014web}. These works also 
required that the token not vary during the 
crawl. In contrast, Fouad et al. did not put a lifetime limit on cookies, 
arguing that trackers can easily link short-lived cookies on their 
servers~\cite{fouad2018missed}. We improve on 
prior work for discarding session IDs by immediately repeating every crawl step 
using a crawler that mimics one previous user. We only assume a token is a 
session ID if it differs across 
these two crawls. This technique allows us to include the 16\% of
\ids{} instances that we would have discarded if CrumbCruncher had used a 90 
day minimum lifetime --- see 
Section~\ref{sec:identifying_uids_static} for more details. 

A further difference between CrumbCruncher and prior work is in how we 
determine if tokens are ``the same'' across users. Some previous work used the 
Ratcliff/Obershelp 
algorithm~\cite{ratcliff-obershelp} to compare cookie values and allowed those 
values to differ by 33\%~\cite{koop2020depth, englehardt2016online, 
acar2014web}, 45\%~\cite{englehardt2015cookies}, or by an unspecified 
amount~\cite{urban2018unwanted}, while still treating the cookies as ``the 
same.'' We chose to 
discard tokens as non-UIDs only when they are entirely identical 
across different users, because we wished to be  
unambiguous about why we had discarded a particular potential UID. Some 
previous work also required cookie lengths to remain the same across 
crawls~\cite{acar2014web, englehardt2015cookies, urban2018unwanted} 
or to only differ by 25\%~\cite{koop2020depth}, as well as requiring 
cookie lengths to be at least eight characters. We require token 
lengths to be greater than eight characters, but we do not place any 
restrictions on the similarity of token lengths across users. 

\subsection{Related work on cookie syncing}

A related technique to navigational tracking is cookie syncing, 
which has been investigated by multiple 
groups~\cite{urban2018unwanted, acar2014web, papadopoulos2019cookie, 
englehardt2016online, papadopoulos2018exclusive, urban2020measuring}. 
Cookie syncing is not the same as navigational tracking, because it 
does not allow third parties to share a UID across top level sites 
when partitioned storage is in use. Instead, cookie syncing allows 
third parties on the same site to share a UID with each other.

\subsection{Other related work}

Trackers may circumvent partitioned storage protections using techniques that 
do not rely on navigational tracking, such as CNAME 
cloaking~\cite{dimova2021cname, dao2020characterizing} or browser 
fingerprinting~\cite{eckersley2010panopticlick, nikiforakis2013fingerprinting, 
Iqbal2021FingerprintingTF}. 

CNAME cloaking is the procedure of mapping a website subdomain to a 
third party domain using a DNS CNAME record. This technique allows 
trackers to share their first party cookies, because the browser is 
tricked into attaching cookies from the original website's subdomain 
rather than the third party domain the subdomain redirects 
to~\cite{dimova2021cname}. 
Trackers can access session cookies, even those belonging to 
financial institutions, using this technique~\cite{aliyeva2021oversharing, 
ren2021analysis}. 

Browser fingerprinting is another technique used by trackers to 
circumvent partitioned storage and track users across websites. 
Browser fingerprinting allows a tracker to use features of a user's 
browser such as window size, installed fonts, supported codecs, and 
more to create a unique ``fingerprint'' of that user that can
function as (or generate) a UID~\cite{eckersley2010panopticlick}. A 2013 study 
crawled 20 
pages for each of the Alexa top 10,000 sites and found that 40 performed 
browser fingerprinting~\cite{nikiforakis2013fingerprinting}. 
A more recent study improved 
detection of fingerprinting code by 
using machine learning~\cite{Iqbal2021FingerprintingTF}. They then 
measured the Alexa top 100,000 sites and found that 10 percent of 
them perform fingerprinting. They find fingerprinting is more common 
with popular sites, as almost 25\% of
the Alexa top 10,000 sites perform fingerprinting.





\section{Conclusion}
\label{sec:conclusion}

In this work, we present the first systematic study of \ids{}, a 
technique that allows trackers to evade browsers' protections against 
cross-website tracking. We find that \ids{} is present across 8.1\% of the 
navigations paths we observed.
We publish a list of the entities that participate in \ids{}, and classify 
these entities according to their behavior and purposes. 
Our findings can be used by browsers to improve protections against 
\ids{}.

Understanding the scope of \ids{}, and the techniques by which 
it is conducted, is important to continue improving privacy on the Web. 
Browsers are increasingly (though not yet universally) trying to protect their
users from being tracked. Understanding how trackers are circumventing
new browser privacy protections is important, to make sure privacy improvements
are not lost as quickly as they're gained.

\bibliographystyle{ACM-Reference-Format}
\bibliography{references}

\end{document}